\documentclass[reprint, notitlepage, secnumarabic,amssymb, aps, prl, figure, superscriptaddress, floatfix] {revtex4-2}
\usepackage{graphicx}
\usepackage{amsmath}
\usepackage{subfigure}
\usepackage{braket}
\usepackage{ulem}
\usepackage{hyperref}
\hypersetup{
    colorlinks=true,
    linkcolor=blue,
    filecolor=red,      
    urlcolor=blue,
    citecolor=magenta
}
\usepackage[version=4]{mhchem}
\usepackage{ragged2e}

\newcommand{\etal}{\textit{et al.}}
\usepackage[usenames, dvipsnames]{color}
\definecolor{mypink}{RGB}{219, 48, 122}
\usepackage{xr}
\usepackage{pdfpages}
\usepackage{pgffor}
\usepackage[pagewise]{lineno}

\makeatletter
\AtBeginDocument{\let\LS@rot\@undefined}
\makeatother

\begin{document}

\title{Going for Gold(-Standard): Attaining Coupled Cluster Accuracy in Oxide-Supported Nanoclusters}

\author{Benjamin X. Shi}
\affiliation{Yusuf Hamied Department of Chemistry, University of Cambridge, Lensﬁeld Road, Cambridge, CB2 1EW, United Kingdom}

\author{David J. Wales}
\affiliation{Yusuf Hamied Department of Chemistry, University of Cambridge, Lensﬁeld Road, Cambridge, CB2 1EW, United Kingdom}

\author{Angelos Michaelides}
\email{am452@cam.ac.uk}
\affiliation{Yusuf Hamied Department of Chemistry, University of Cambridge, Lensﬁeld Road, Cambridge, CB2 1EW, United Kingdom}

\author{Chang Woo Myung}
\email{cwmyung@skku.edu}
\affiliation{Department of Energy Science, Sungkyunkwan University, Seobu-ro 2066, Suwon, 16419, Korea}
\affiliation{Yusuf Hamied Department of Chemistry, University of Cambridge, Lensﬁeld Road, Cambridge, CB2 1EW, United Kingdom}

\date{\today}

\begin{abstract}
The structure of oxide-supported metal nanoclusters plays an essential role in their sharply enhanced catalytic activity over bulk metals. 
Simulations provide the atomic-scale resolution needed to understand these systems. 
However, the sensitive mix of metal-metal and metal-support interactions which govern their structure puts stringent requirements on the method used, requiring going beyond standard density functional theory (DFT).
The method of choice is coupled cluster theory [specifically CCSD(T)], but its computational cost has so far prevented applications.
In this work, we showcase two approaches to make CCSD(T) accuracy readily achievable in oxide-supported nanoclusters. 
First, we leverage the SKZCAM protocol to provide the first benchmarks of oxide-supported nanoclusters, revealing that it is specifically metal-metal interactions that are challenging to capture with DFT.
Second, we propose a CCSD(T) correction ($\Delta$CC) to the metal-metal interaction errors in DFT, reaching comparable accuracy to the SKZCAM protocol at significantly lower cost. 
This forges a path towards studying larger systems at reliable accuracy, which we highlight by identifying a ground state structure in agreement with experiments for \ce{Au20} on MgO; a challenging system where DFT models have yielded conflicting predictions.
\end{abstract}

\maketitle

\section{Introduction}

Metal nanoclusters stabilized on metal-oxide surfaces (i.e.,\ oxide-supported nanoclusters) form catalysts that underpin a wide array of industrial applications, encompassing water splitting~\cite{harzandiImmiscibleBimetalSingleatoms2020,tiwariMulticomponentElectrocatalystUltralow2018}, \ce{CO2} reduction~\cite{jangElectrocatalyticCO2Reduction2022,yuanSituManipulationActive2021a}, the reverse water-gas shift reaction~\cite{pahijaExperimentalComputationalSynergistic2022} and beyond~\cite{rongSyntheticStrategiesSupported2020a,koDirectPropyleneEpoxidation2022a,ringeImportanceChargeTransfer2023}.
Their enhanced and tunable catalytic properties~\cite{campbellEnthalpiesEntropiesAdsorption2013}, combined with high surface-to-volume ratio, make them particularly appealing candidates for next-generation catalysts that address the urgent climate crisis~\cite{campbellEffectSizeDependentNanoparticle2002,beniyaCOOxidationActivity2020b,vandeelenControlMetalsupportInteractions2019a,zandkarimiSurfacesupportedClusterCatalysis2019a,chattotSurfaceDistortionUnifying2018a,sunSolidtoliquidPhaseTransitions2019}.
At small (${<}4\,$nm) sizes, the structure that the metal nanocluster adopts on the metal-oxide support plays a central role in tailoring its catalytic activity~\cite{vandeelenControlMetalsupportInteractions2019a}.
For example, catalytic reaction rates can change significantly across different structures of the same nanocluster - attributed to the appearance of various under-coordinated sites~\cite{norskovNatureActiveSite2008}. %
In fact, under normal reaction temperatures, the metal clusters are fluxional~\cite{zandkarimiSurfacesupportedClusterCatalysis2019a,chattotSurfaceDistortionUnifying2018a,sunSolidtoliquidPhaseTransitions2019,nianAtomicScaleDynamicInteraction2020a,sunSurfacereactionInducedStructural2020a}, existing as several low-energy structures that evolve during the course of the reaction.

It is essential that the relative energies, and thus stabilities, between the low-energy structures of an oxide-supported nanocluster can be determined accurately.
This challenge is most commonly tackled with computational simulations~\cite{ricciBondingTrendsDimensionality2006a,sterrerCrossoverThreeDimensionalTwoDimensional2007,vilhelmsenSystematicStudyAu62012} using density functional theory (DFT).
To date, DFT has provided valuable insights into the low-energy structures of nearly every pairing of elemental metal nanocluster and metal-oxide surface~\cite{remediakisCOOxidationRutileSupported2005a, dingGrowthPatternAun2015,molinaActiveRoleOxide2003,yoonChargingEffectsBonding2005c,huberOxidationMagnesiasupportedPdclusters2006,qiaoSingleatomCatalysisCO2011}.
However, these systems encompass complex metal-metal interactions (within the nanocluster) and metal-support interactions (i.e.\ charge-transfer and van der Waals attraction between the nanocluster and oxide surface), which can prove challenging to model with DFT~\cite{tosoniOxideSupportedGold2019}.
In particular, predictions on the relative stabilities of low-energy structures have been shown to be highly sensitive to the DFT model~\cite{kinaciUnravelingPlanarGlobularTransition2016} [i.e.\ exchange-correlation (XC) functional and dispersion correction] employed, particularly around the nanocluster 2D-3D transition size~\cite{guedes-sobrinhoEffectDifferentEnergy2022,bulusuEvidenceHollowGolden2006} - the size above which a nanocluster favors a 3D over a 2D planar structure.
For example, the 2D-3D transition size for gas-phase gold (Au$_N$) nanoclusters has been predicted by various DFT models to be anywhere from \ce{Au7}~\cite{hakkinenGoldClustersAuN2000} to \ce{Au15}~\cite{xiaoPlanarThreedimensionalStructural2004}.
Crucially, it is not possible to know \textit{a priori}, without suitable benchmark references~\cite{johansson2D3DTransitionGold2008,baekBenchmarkStudyDensity2017b,sunParametrizationPM7Semiempirical2022,paz-borbonAuNClusters162012}, which DFT model to trust.

\begin{figure*} [t]
\includegraphics[width=6.75in]{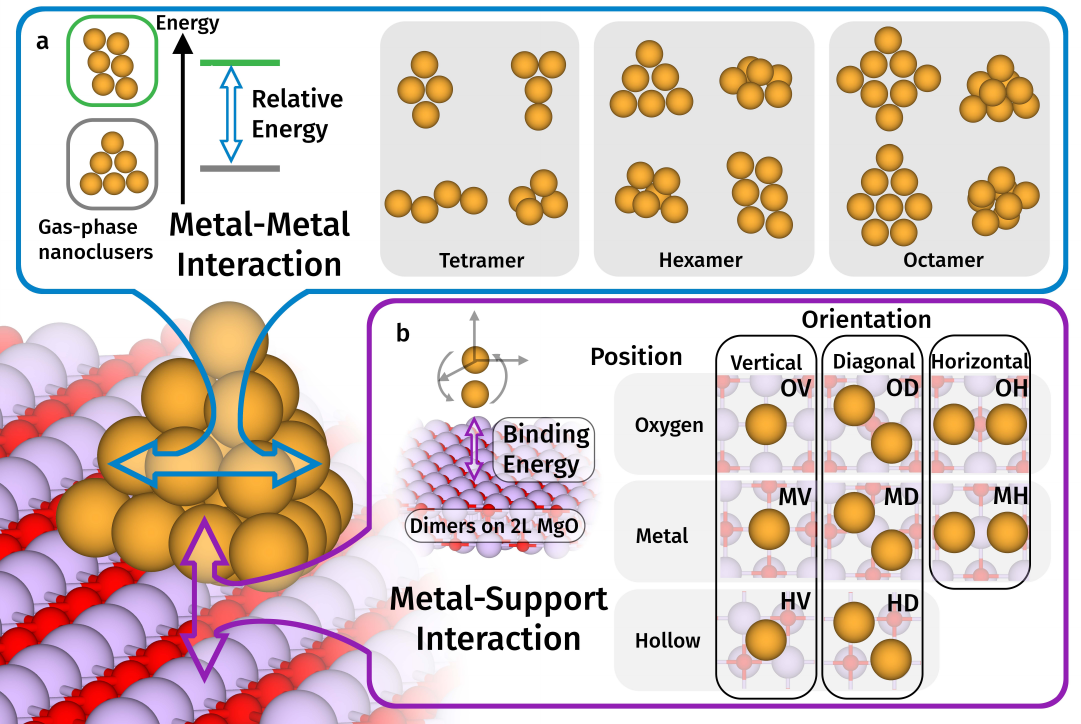}
\caption{Schematic of the set of structures used to benchmark various DFT models. (a) We investigate the metal-metal interactions by comparing relative energy ordering of tetramers, hexamers and octamers in the gas-phase against LNO-CCSD(T). (b) Metal-support interactions are investigated by calculating the binding energy of coinage metal dimers (in 8 different adsorbed configurations) on a two-layer MgO slab.}
\label{fig:test-set}
\end{figure*}

The presence of a support material has been shown to further exacerbate the discrepancies between DFT models on the energetics of the nanocluster structure.
Many examples exist within the literature~\cite{srivastavaDensityFunctionalStudy2012a,yuanStructuresCatalyticProperties2014,engelInfluenceSupportMaterials2019a,tengDFTStudyStructures2012,vilhelmsenSystematicStudyAu62012} and we focus particularly on Au$_N$ on MgO -- the most studied system by far.
For example, the 2D-3D transition size is predicted to be between \ce{Au16} and \ce{Au20} for the PBE functional~\cite{ferrandoStructuresSmallAu2011a,ricciBondingTrendsDimensionality2006a} while it is smaller than \ce{Au11} for LDA~\cite{ferrandoStructuresSmallAu2011a}. 
There are even discrepancies in the global minimum structure prediction in the planar 2D regime for PBE, with some suggesting planar 2D structures that lie either perpendicular~\cite{ferrandoStructuresSmallAu2011a,paz-borbonAuNClusters162012,buendiaComparativeStudyAumRhn2016} or parallel to the surface~\cite{ricciBondingTrendsDimensionality2006a,stamatakisMultiscaleModelingReveals2012a,gaoInfluencesMgO0012020,engelInfluenceSupportMaterials2019a}.
By augmenting the PBE functional with the D3 dispersion correction, Engel \etal{}~\cite{engelInfluenceSupportMaterials2019a} have actually found that planar parallel structures exist up until \ce{Au19} - their largest studied nanocluster - and we show it extends up to \ce{Au20} as well in this study. 
PBE-D3 is perhaps the most common DFT model used in surface chemistry but alarmingly, these predictions by Engel \etal{} are in stark contrast to the 3D tetrahedral structure (see Fig.~\ref{fig:test-set}) predicted by other studies~\cite{ferrandoStructuresSmallAu2011a,ricciBondingTrendsDimensionality2006a,mammenInducingWettingMorphologies2018} as well as strong experimental evidence~\cite{hardingControlManipulationGold2009,wangDirectAtomicImaging2012,liUnravelingAtomicStructure2020a}.

There is, thus, an urgent need to study oxide-supported metal nanoclusters with a benchmark method, especially in the absence of experiments.
To this end, methods from correlated wave-function theory (cWFT) offer a systematic route to solving the many-body electron problem.
Of these, the coupled cluster (CC) family~\cite{bartlettCoupledclusterTheoryQuantum2007} is very popular, reaching the exact solution by progressively expanding a reference wave-function with particle-hole excitation operators of arbitrary order (up to all $n$ electrons in a system).
CC truncated at the single, double and perturbative triple particle-hole excitation operator level [i.e.\ CCSD(T)~\cite{raghavachariFifthorderPerturbationComparison1989}] has been shown to provide reliable predictions for a wide array of systems~\cite{kartonW4TheoryComputational2006}.
Unfortunately, the steep scaling of its computational complexity with system size [$\mathcal{O} \left (n^7 \right )$] relative to DFT [$\mathcal{O} \left (n^3 \right )$] makes it highly challenging to apply.
This problem is particularly acute for oxide-supported nanoclusters because it involves: (i) a surface; and (ii) heavy transition-metals in the nanocluster, which naturally lead to large system sizes.
Thus, to our knowledge, no study of these systems has so far incorporated CCSD(T) or even the lower-level second-order M\o{}ller Plesset perturbation theory (MP2), with only one such study using the random phase approximation (RPA) on small (Au$_2$ to Au$_6$) nanoclusters~\cite{paz-borbonAuNClusters162012}.

Without the availability of benchmark methods, it has not been possible to identify robust DFT models for oxide-supported nanoclusters.
Indeed, we show in this work that the commonly used PBE-D3 DFT model suffers systematic errors, which makes its application to these systems inadvisable.
Fortunately, several developments in recent years have now set the stage for oxide-supported nanoclusters to be studied with CCSD(T). 
Advances in localized orbital variants of CCSD(T) (e.g.\ LNO-CCSD(T)~\cite{nagyApproachingBasisSet2019}, DLPNO-CCSD(T)~\cite{riplingerEfficientLinearScaling2013}, PNO-LCCSD(T)~\cite{maExplicitlyCorrelatedLocal2018}, etc.) that achieve linear scaling (under ideal circumstances) now significantly broaden the scope of system sizes that can be investigated. 
These developments have been largely confined to molecular (gas-phase) systems as their extension to solids (and surfaces) is not straightforward~\cite{kubasSurfaceAdsorptionEnergetics2016,sauerInitioCalculationsMolecule2019b}. 
Recently, Shi, Kapil, Zen, Chen, Alavi, and Michaelides developed the  SKZCAM protocol~\cite{shiGeneralEmbeddedCluster2022b} to tackle this challenge.
In combination with LNO-CCSD(T), it has demonstrated reliable predictions for properties ranging from adsorption~\cite{shiManyBodyMethodsSurface2023a} to vacancy formation energies~\cite{shiGeneralEmbeddedCluster2022b}.

In this work, we showcase two approaches - the SKZCAM protocol and DFT+$\Delta$CC - to achieve CCSD(T) accuracy in oxide-supported nanoclusters for the first time.
With these two methods, we provide reliable insights into the structure of coinage metal (Au, Ag and Cu) clusters supported on MgO - the prototypical system with highly sought-after catalytic capabilities~\cite{luoReactivityMetalClusters2016}.
The SKZCAM protocol was used to benchmark small ($N \leq 4$) clusters to assess the accuracy of a wide range of DFT models, encompassing a broad set of exchange-correlation functionals (across Jacob's ladder~\cite{perdewJacobLadderDensity2001a}) and dispersion corrections~\cite{klimesPerspectiveAdvancesChallenges2012c}. 
We demonstrate that none of the examined DFT models can represent these systems with high accuracy, finding that while metal-support interactions could be reliably predicted by one DFT model (PBE0-TS/HI), there were large errors in the metal-metal interactions of all DFT models. 
We show that errors in the metal-metal interactions can be fixed through a $\Delta$CC~\cite{sauerInitioCalculationsMolecule2019b} correction that confines the CCSD(T) calculations to only the metal nanocluster.
The resulting DFT+$\Delta$CC estimates (correcting PBE0-TS/HI) were verified to reach the same accuracy as the SKZCAM protocol estimates at much lower cost.
Our confidence in its accuracy opens the door towards studying larger oxide-supported nanoclusters, such as the (aforementioned) disputed geometry of Au\textsubscript{20} on MgO. %
In particular, we are able to demonstrate definitive agreement with experimental evidence that the 3D tetrahedral structure is the most stable for this system.

\section{Methods}

The quantity that can be computed to establish the stability of a supported nanocluster structure is its (electronic) total energy; the lower (i.e., more negative) its energy, the more stable the nanocluster.
In this study, we specifically compare the relative energies $\Delta_\textrm{rel.}$ between nanocluster structures.
For a nanocluster structure $Y$ (either in the gas-phase or on the oxide surface), its relative energy to the ground-state structure $X$ is given as:
\begin{equation}
    \Delta_\textrm{rel.}^{Y} = E_\textrm{total}[Y] - E_\textrm{total}[X],
\end{equation}
where $E_\textrm{total}[..]$ is the total energy of the structure enclosed in the square brackets.

The relative energy between these structures is dictated by a combination of the metal-metal interactions within the nanocluster and the metal-support interactions between the nanocluster and oxide support.
We directly probe the metal-metal interactions by comparing $\Delta_\textrm{rel.}$ for unsupported nanocluster structures in the gas-phase.
Metal-support interactions can be probed by computing the binding energy of a supported nanocluster to the oxide surface:
\begin{equation}
    E_\textrm{bind} = E[\textrm{NC+Ox}] - E[\textrm{NC}] - E[\textrm{Ox}],
\end{equation}
where NC, Ox and NC+Ox represent the nanocluster, oxide surface and the oxide-supported nanocluster respectively.
The NC and Ox structures have their geometry frozen, fixed from the optimized NC+Ox system.

\subsection{Computational details for DFT}
Periodic DFT with the Vienna \textit{Ab-initio} Simulation Package (VASP)~\cite{kresseEfficiencyAbinitioTotal1996a,kresseEfficientIterativeSchemes1996a} was used generate the geometries to calculate the CCSD(T) relative and binding energy references.
Subsequently, we benchmarked a broad set of 42 DFT exchange-correlation functionals and empirical vdW corrections on  (see Sec.~\ref{si-sec:dft_methods} of Supporting Information). 
To accurately model the MgO (001) surface, a $5{\times}5$ supercell with 50 atoms in each layer was used, where a two layer (2L) thick slab was used for the adsorption of coinage metal dimers and tetramers while a 4L slab was used for the adsorption of Au\textsubscript{20}.
A total of $23.4\,$\AA{} of vacuum was placed along the surface normal direction. 
The rev-vdW-DF2 DFT model was used to perform all geometry (and global) optimizations, chosen due to its good performance in reproducing the lattice parameter and coinage metal dimer binding energies to MgO (in Secs.~\ref{si-sec:dft_methods} and~\ref{si-sec:dimer_benchmark} of the Supporting Information respectively).
The calculations performed with a 2L surface had all layers fixed to the bulk while those employing a 4L surface were relaxed to a force threshold of $0.01\,$eV\AA{}$^{-1}$ (with the bottom two layers fixed).
Standard projector-augmented wave potentials, a $520\,$eV energy cutoff, dipole corrections and a $\Gamma$ $k$-point mesh were used. 
For the DFT models that incorporated the many-body dispersion schemes~\cite{tkatchenkoAccurateEfficientMethod2012}, a (4 $\times$ 4 $\times$ 1) k-point mesh was used. 
Global optimizations, such as the evolutionary algorithm~\cite{bisboEfficientGlobalStructure2020} coupled with the sparse Gaussian process regression (SGPR) machine learning potential~\cite{hajibabaeiSparseGaussianProcess2021}, and basin-hopping~\cite{walesGlobalOptimizationBasinHopping1997b}, were used to find low energy gas phase structures (\ce{M4}, \ce{M6}, \ce{M8}) and nanocluster structures (\ce{M4} and \ce{Au20}) on MgO, where M is Cu, Ag and Au. 

\subsection{Computational details for CCSD(T)}

The correlated wave-function theory (cWFT) calculations [i.e.\ MP2, CCSD, CCSD(T), CCSDT(Q)] were performed in MRCC~\cite{kallayMRCCProgramSystem2020}.
All calculations (aside from the validations in Sec.~\ref{si-sec:cc_validity} of the Supporting Information)  utilized the localized natural orbital (LNO) approximation with the LNO-CCSD(T), LNO-CCSD and local MP2 (LMP2) implementations of Nagy \etal{}~\cite{nagyIntegralDirectLinearScalingSecondOrder2016,nagyOptimizationLinearScalingLocal2018}. 
The correlation treatment explicitly includes subvalence electrons~\cite{bistoniTreatingSubvalenceCorrelation2017} on the Mg (2s and 2p), Cu (3s and 3p), Ag (4s and 4p) and Au (5s and 5p) metal atoms, alongside corresponding cc-pwCV$n$Z~\cite{petersonAccurateCorrelationConsistent2002,balabanovSystematicallyConvergentBasis2005} basis sets. 
The O atoms used the aug-cc-pV$n$Z~\cite{kendallElectronAffinitiesFirst1992} basis set. 
The def2/JK~\cite{weigendHartreeFockExchange2008a} basis set was used as the auxiliary basis function for all atoms. 
The corresponding RI auxiliary basis sets~\cite{weigendRIMP2OptimizedAuxiliary1998,hellwegOptimizedAccurateAuxiliary2007} were used for the local cWFT calculations, but the automatic auxiliary basis functions of Stoychev \etal{}~\cite{stoychevAutomaticGenerationAuxiliary2017,lehtolaStraightforwardAccurateAutomatic2021} were generated for the Mg and coinage metal basis sets.
For the coinage metals, we used the MCDHF effective core potentials from the Stuttgart/Cologne group~\cite{figgenEnergyconsistentPseudopotentialsGroup2005}, which incorporate scalar-relativistic and spin–orbit effects from the core electrons.
Complete basis set (CBS) extrapolation parameters for the TZ and QZ pair, CBS(TZ/QZ), taken from Neese and Valeev~\cite{neeseRevisitingAtomicNatural2011}, were used for the Hartree-Fock (HF) and correlation energy components of the cWFT total energy. 
The binding energies of Au dimers on the MgO surface were computed with basis-set superposition error (BSSE) corrections~\cite{boysCalculationSmallMolecular1970}. 
On the other hand, relative energies did not require BSSE corrections, as we show in Sec.~\ref{si-sec:bind_ene_calc} of the Supporting Information.
The binding energy calculations of the oxide-supported nanoclusters were performed with an electrostatic embedded cluster approach whereby the point charge environment was constructed using py-ChemShell~\cite{luOpenSourcePythonBasedRedevelopment2019a}, with the quantum clusters generated using the SKZCAM protocol~\cite{shiGeneralEmbeddedCluster2022b}.

\section{Results and Discussion}

\subsection{An accurate dataset to benchmark oxide-supported nanoclusters}

The CCSD(T)-quality dataset within this work (Fig.~\ref{fig:test-set}) has been developed to assess DFT models for their performance in treating the metal-metal and metal-support interactions separately; these are the key contributions which dictate the lowest energy geometry of a supported nanocluster.
The metal-metal interactions were investigated (Fig.~\ref{fig:test-set}a) by comparing the relative energy between four Au$_N$, Ag$_N$ and Cu$_N$ ($N=4,6,8$) gas-phase structures, as determined through a global optimization scheme (described in Sec.~\ref{si-sec:geom_global_opt} of the Supplementary Information).
Metal-support interactions were probed by comparing the binding energy of coinage metal dimers across several orientations and positions above the MgO surface (Fig.~\ref{fig:test-set}b).
Dimers were chosen as they allow us to systematically study the performance of various DFT models across specific binding sites.

Attaining a CCSD(T)-level of accuracy on oxide-supported nanoclusters was only made possible thanks to the combination of the LNO approximation to CCSD(T) and SKZCAM protocol to approximate the surface.
In Sec.~\ref{si-sec:cc_validity} of the Supporting Information, we have performed extensive tests to confirm that the error introduced by these approximations is under control.
First, we have benchmarked LNO-CCSD(T) against canonical CCSD(T), finding mean absolute errors of $51\,$meV in the relative energies of coinage metal tetramers.
Second, for the embedded clusters used to represent the MgO surface in the SKZCAM protocol, we have shown that errors are within $30\,$meV w.r.t.\ the bulk (infinite surface size) limit.
As summarized in Sec.~\ref{si-sec:error_bars} of the Supporting Information, these errors are close to chemical accuracy of $43\,$meV ($1\,$kcal/mol) -- the typical target needed for reliable predictions of thermodynamic properties.
Furthermore, we also confirm that CCSD(T) serves as an appropriate benchmark for the relative energies of the coinage metal tetramers by validating it against CCSDT(Q) -- CC truncated to the perturbative quadruples level -- finding a mean absolute deviation (MAD) of only $13\,$meV.
Additional multireference diagnostics (in Sec.~\ref{si-sec:cc_even_validity} of the Supporting Information) for larger clusters also support CCSD(T) as appropriate for the closed-shell even-numbered coinage metal nanoclusters investigated in this study.

The accuracy of our CCSD(T) benchmarks of the Au gas-phase nanoclusters in Fig.~\ref{fig:test-set}a can be further validated against the previous literature.
In Sec.~\ref{si-sec:lit_copare} of the Supporting Information, we have compared our LNO-CCSD(T) predictions to these previous estimates, finding that deviations are less than $120\,$meV (${\sim}3\,$ kcal/mol).
These discrepancies likely arise from differing nanocluster structures (optimized with different DFT models or even cWFT methods), or inadequacies in the electronic structure treatment.
The use of LNO-CCSD(T) here has allowed this work to go to larger basis sets (triple and quadruple-zeta extrapolated) than before, while also explicitly correlating semicore 5s and 5p electrons in Au.
The only main source of error in our relative energy estimates is in employing the LNO approximation, which contributes around a $50\,$meV error (see Sec.~\ref{si-sec:error_bars} of the Supporting Information).
This level of accuracy was achieved while requiring significantly less resources and as we show later, it allows us to go to systems such as Au\textsubscript{20}; a system size far outside the reach of canonical CCSD(T).
In addition, our LNO-CCSD(T) calculations demonstrate that the ground state of Au\textsubscript{8} has a 2D planar $D_{4h}$ symmetry, in agreement with recent experimental observations~\cite{grueneFarIRSpectraSmall2014}.
This system has been highly challenging to treat with CCSD(T), where predictions have been split between either 2D~\cite{diefenbachSpatialStructureAu82006,hanStructureAu8Planar2006,olsonIsomersAu82007} and 3D~\cite{olsonWhereDoesPlanartoNonplanar2005} ground state structures.
It has been shown in Ref.~\citenum{hansenCommunicationDeterminingLowestenergy2013} that predicting the correct geometry necessitates converged basis sets and inclusion of semicore electrons, which we explicitly treat here.

\subsection{Insights into the performance of DFT models}

With our CCSD(T)-quality dataset validated, we can now assess how DFT models (XC functional and dispersion correction) perform for oxide-supported nanoclusters.
It would be impossible to provide an exhaustive assessment of all DFT models and we focus only on some of the most widely-used models for surface chemistry.
Specifically, we have compared 42 DFT models (see Sec.~\ref{si-sec:gas_phase_nanocluster_energies} of the Supporting Information), which span Jacob's ladder, starting from generalized gradient approximations [GGAs] (PBE~\cite{perdewGeneralizedGradientApproximation1996c}, PBEsol~\cite{perdewRestoringDensityGradientExpansion2008a}, revPBE~\cite{zhangCommentGeneralizedGradient1998}) to van der Waals (vdW)-inclusive (vdW-DF~\cite{dionVanWaalsDensity2004}, vdW-DF2~\cite{leeHigheraccuracyVanWaals2010}, rev-vdW-DF2~\cite{hamadaVanWaalsDensity2014} and optB86b-vdW~\cite{klimesVanWaalsDensity2011a}), meta-GGA (r\textsuperscript{2}SCAN~\cite{furnessAccurateNumericallyEfficient2020a,sunStronglyConstrainedAppropriately2015b}, TPSS~\cite{taoClimbingDensityFunctional2003} and M06L~\cite{zhaoNewLocalDensity2006}), hybrid (HSE06~\cite{krukauInfluenceExchangeScreening2006}, PBE0~\cite{perdewRationaleMixingExact1996,adamoReliableDensityFunctional1999c} and B3LYP~\cite{stephensInitioCalculationVibrational1994}) and double-hybrid (B2PLYP~\cite{grimmeSemiempiricalHybridDensity2006}) exchange-correlation functionals.
This selection of functionals was combined with the D2~\cite{grimmeSemiempiricalGGAtypeDensity2006}, D2[Ne]~\cite{tosoniAccurateQuantumChemical2010a}, D3~\cite{grimmeConsistentAccurateInitio2010}, D4~\cite{caldeweyherGenerallyApplicableAtomiccharge2019}, MBD~\cite{tkatchenkoAccurateEfficientMethod2012}, MBD/HI, MBD/FI~\cite{gouldFractionallyIonicApproach2016}, TS~\cite{tkatchenkoAccurateMolecularVan2009}, TS/HI~\cite{buckoImprovedDensityDependent2013}, rVV10~\cite{ningWorkhorseMinimallyEmpirical2022} and DDsC~\cite{steinmannGeneralizedgradientApproximationExchange2011} dispersion corrections (discussed in Ref.~\citenum{r.rehakIncludingDispersionDensity2020})

\begin{figure}[htp!]
\includegraphics[width=3.375in]{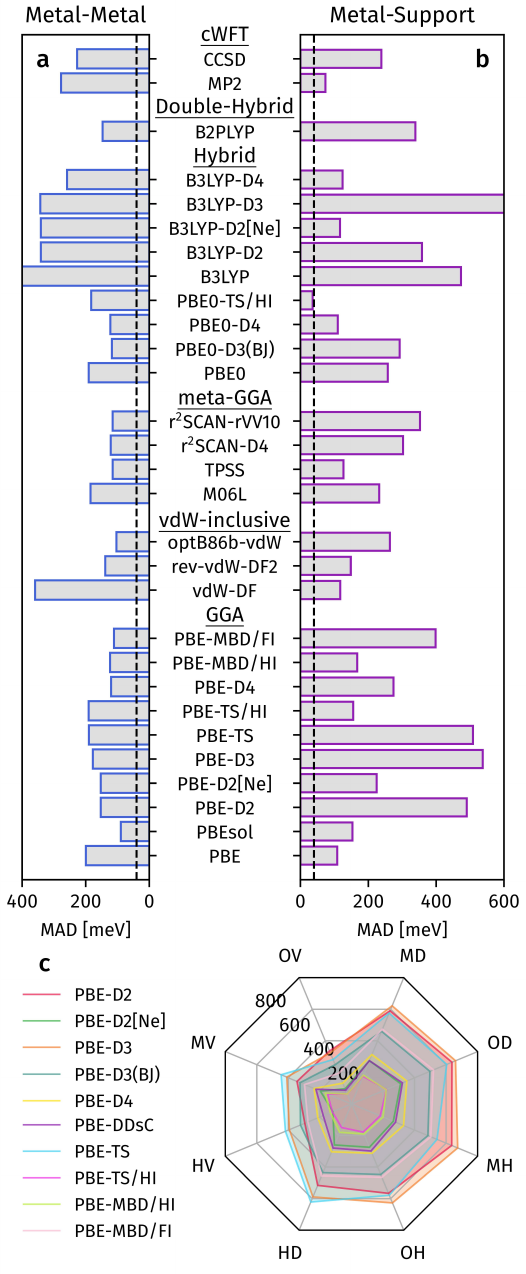}
\caption{A comparison of several DFT models against CCSD(T) for (a) metal-metal and (b) metal-support interactions, with the dotted line indicating chemical accuracy. For the metal-support interactions, we compare the various dispersion corrections for PBE in (c). The 8 different adsorption configurations considered in (c) are shown in Fig.~\ref{fig:test-set}b.}
\label{fig:dft-benchmark}
\end{figure}

In Fig.~\ref{fig:dft-benchmark}a, we highlight the mean absolute deviation (MAD) for a subset of DFT models to CCSD(T) in predicting the relative energies of the gas-phase tetramer, hexamer and octamer structures across all three coinage metals.
Overall, we come to the conclusion that no DFT model can achieve the sought-after chemical accuracy for metal-metal interactions. 
The DFT model that provides the lowest MAD of $89\,$meV ($\sim{}2\,$kcal/mol) is PBEsol, while B3LYP was the worst performing DFT model out of all that were studied (see Sec.~\ref{si-sec:gas_phase_nanocluster_energies} of the Supporting Information) at an MAD of $406\,$meV.
The poor performance of the latter functional is in agreement with previous assessments~\cite{baekBenchmarkStudyDensity2017b}. 
In general, we find that all XC functionals show improvements when dispersion corrections are added, echoing the conclusions of several recent works~\cite{goldsmithTwotothreeDimensionalTransition2019a,luna-valenzuelaEffectsVanWaals2021}.
On the other hand, there are no systematic trends moving up the rungs of Jacob's ladder, with meta-GGAs performing better than many hybrid models and even the double-hybrid B2PLYP.

\begin{figure*}[ht!]
\includegraphics[width=6in]{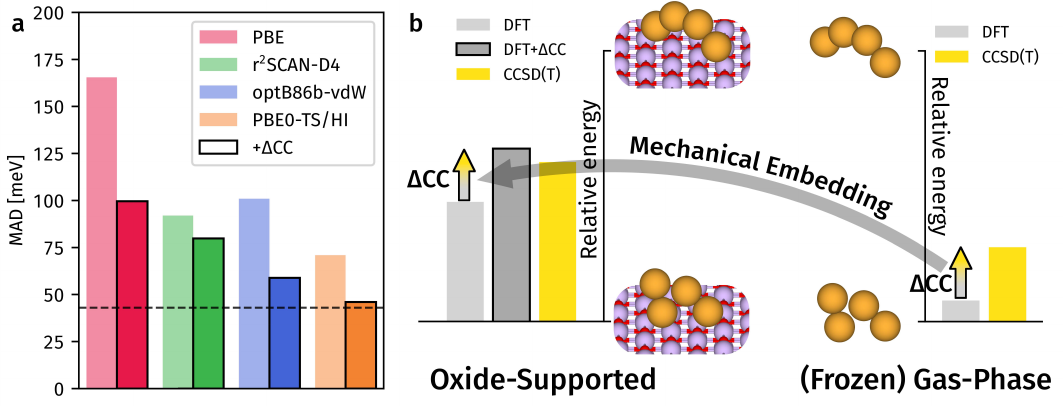}
\caption{(a) The mean absolute deviation (MAD) of the PBE, r\textsuperscript{2}SCAN-D4, optB86b-vdW and PBE0-TS/HI DFT models to LNO-CCSD(T) (using the SKZCAM protocol) for the relative energies of coinage metal tetramers on MgO (see Fig.~\ref{si-fig:oxide-support_geometries} of the Supporting Information). The improvements for each DFT model with the $\Delta$CC correction are also shown. (b) This $\Delta$CC correction works by extracting the (frozen) nanoclusters from the oxide-supported nanocluster and rectifying the discrepancies in their energy ordering (in the gas-phase) at the DFT level, up to the CCSD(T) level (see text).}
\label{fig:dft-cc-schematic}
\end{figure*}

Unlike metal-metal interactions, metal-support interactions have not been benchmarked before.
However, the SKZCAM protocol now enables their treatment at CCSD(T)-level in this study.
While all DFT models in this work and previous work~\cite{barcaroInteractionCoinageMetal2005,caballeroStructuralElectronicProperties2008,fuenteFormationAg2Au22009} agree that the lowest energy structure of a coinage metal dimer is in the vertical position on top of the O atom (i.e.\ OV in Fig.~\ref{fig:test-set}b), there has been large noted discrepancies in the strength of this binding relative to different configurations~\cite{fuenteFormationAg2Au22009}.
For the same set of DFT models as the metal-metal interactions, we have assessed the performance for metal-support interactions (i.e.\ the binding energy of various dimer configurations) in Fig.~\ref{fig:dft-benchmark}b. 
Notably, we observe a wider range of errors in metal-support interactions, starting from as small as $36\,$meV to as high as $600\,$meV. 
The best-performing DFT model, achieving the coveted chemical accuracy, is PBE0-TS/HI. 
Its good performance can be understood to some extent.
First, the underlying PBE0 functional predicts excellent band gaps for MgO~\cite{dittmerAccurateBandGap2019}, so charge-transfer effects will be approximated well.
Second, we have compared a large set of dispersion corrections (in Fig.~\ref{fig:dft-benchmark}c) to the PBE functional and found that TS/HI was the best performing across all 8 dimer configurations (i.e.\ binding sites).
As we discuss below, its excellent performance arises because it does not suffer from an overbinding in ionic systems that afflicts many dispersion corrections.

We do not find any clear trends between different DFT models for the metal-support interactions.
For example, along Jacob's ladder, MADs can vary by over $400\,$meV within a single rung.
Several DFT models from lower rungs, especially vdW-inclusive ones, can perform better than the sophisticated B2PLYP double-hybrid functional or even CCSD - a cWFT method.
Dispersion corrections do not always give improvements over the underlying XC functional, especially for PBE, where all dispersion corrections give higher MADs.
In fact, the dispersion corrections cause an overbinding for all 8 dimer configurations relative to PBE.
This effect, as shown in Sec.~\ref{si-sec:c6_issue} of the Supporting Information, arises from an overestimated C\textsubscript{6} (dispersion) parameter on the Mg atom in many of the dispersion corrections~\cite{ehrlichSystemDependentDispersionCoefficients2011}.
It explains why PBE-D2[Ne] performs better than PBE-D2, as the former replaces the Mg C\textsubscript{6} D2 parameters with those from neon as it has been recognized that the original D2 parameters highly overestimate this quantity for the Mg cation (and indeed many other cations)~\cite{tosoniAccurateQuantumChemical2010a}.
Similarly, methods that naturally account for electron density changes, such as vdW-inclusive DFT models and dispersion corrections that incorporate iterative Hirshfeld partitioning (HI) do not suffer from this problem~\cite{r.rehakIncludingDispersionDensity2020}.

\subsection{Reaching larger systems at coupled cluster accuracy}

The overall DFT errors in the relative energy ordering of \textit{oxide-supported} nanocluster with different geometries are likely to arise from a description of both metal-metal and metal-support interactions.
In Fig.~\ref{fig:dft-cc-schematic}a, we have computed the relative energy ordering of four tetramer configurations for the same set of DFT models (only a subset is shown) as before and computed the MAD w.r.t.\ LNO-CCSD(T) estimates using the SKZCAM protocol.
We find that PBE0-TS/HI is the best, as a consequence of its excellent performance in metal-support interactions.
However, the error is still relatively significant at around $71\,$meV (${\sim}2\,$kcal/mol).

Our findings in the previous section indicate that the primary remaining errors in PBE0-TS/HI arise from metal-metal interactions.
Consequently, we propose a simple correction, based upon mechanical embedding~\cite{sauerInitioCalculationsMolecule2019b}, which specifically targets these interactions (see Fig.~\ref{fig:dft-cc-schematic}b).
This approach involves extracting the (frozen) nanocluster geometries adsorbed onto the MgO surface and computing only these (now gas-phase) nanoclusters at the (LNO-)CCSD(T) level and with a chosen DFT model.
The errors in the metal-metal interactions can then be corrected by adding the $\Delta$CC correction between CCSD(T) and the DFT model onto the full DFT model calculation on the oxide-supported nanocluster.

In Fig.~\ref{fig:dft-cc-schematic}a, we have highlighted the MADs using this DFT+$\Delta$CC approach for several DFT models.
The $\Delta$CC correction improves the MADs for most DFT models (see Sec.~\ref{si-sec:tetramers_mgo} of the Supporting Information), with several reaching near chemical accuracy to the SKZCAM protocol.
Under the presumption that metal-metal interaction errors are completely remedied, PBE0-TS/HI+$\Delta$CC is expected to perform best.
Indeed, we observe that it is one of the best, and while some other functionals may give better MADs, these improvements are small, within the aforementioned errors from using the SKZCAM protocol and CCSD(T).
In the event that (hybrid) PBE0-TS/HI is not feasible, we recommend opting for the cheaper rev-vdW-DF2+$\Delta$CC, which is the overall best performing DFT model with an MAD of $38\,$meV (see Sec.~\ref{si-sec:tetramers_mgo} of the Supporting Information).

\begin{figure}[htp]
\includegraphics[width=3.37in]{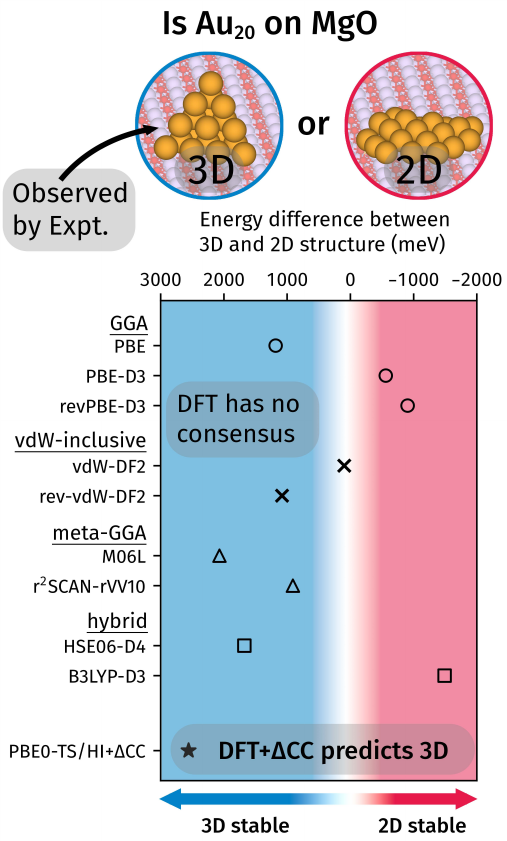}
\caption{Relative energies (eV) of low lying \ce{Au20} nano-clusters on \ce{MgO (100)} at various DFT levels. Depending on the functional used, very different global minima are predicted for \ce{Au20}. Here, a negative energy difference means that the 2D planar structure is more stable than the 3D tetrahedral structure.}
\label{fig:au20-mgo}
\end{figure}

As discussed earlier, the \ce{Au20} on MgO system represents a particularly challenging case for DFT models.
This system size is unfortunately too large for the SKZCAM protocol to tackle.
However, the lower cost of the PBE0-TS/HI+$\Delta$CC approach allows it to be applied to this system to resolve these questions. 
As shown in Fig.~\ref{fig:au20-mgo}, it is challenging for DFT functionals to predict the correct geometry of this system, with some predicting the 3D structure to be stable while others predict a 2D structure.
On the other hand, there is strong evidence that Au$_N$ clusters near $N=20$ take up 3D structures~\cite{sterrerCrossoverThreeDimensionalTwoDimensional2007,giorgioDynamicObservationsAu2008}.
The tetrahedral structure has been observed on the NaCl surface~\cite{liUnravelingAtomicStructure2020a} - an analogous surface and system to MgO - as well as supported carbon~\cite{wangDirectAtomicImaging2012a,liAtomicscaleObservationDynamical2015}.
Importantly, as this 3D shape is maintained at room temperature, it suggests that the static energy difference between the 3D and 2D structures has to be large, of the order of $k_B T$/atom (or $500\,$meV at room temperature).

As seen in Fig.~\ref{fig:au20-mgo}, PBE0-TS/HI+$\Delta$CC predicts the 3D shape to be stable by more than $2500\,$meV, such that we expect it to exist as only the 3D shape on MgO. 
On the other hand, few DFT models are able to reach more than 500 meV, with their (large) discrepancies highlighting the importance of balancing effects such as vdW dispersion and exchange. 
Most notably, PBE-D3 and revPBE-D3 (with zero-damping) predicts a 2D shape to be stable and we see significant changes just by altering the D3 damping used (see Sec.~\ref{si-sec:au20_benchmarks} of the Supporting Information). 
Similarly, while vdW-DF2 suggests 3D and 2D structures are effectively degenerate, rev-vdW-DF2 strongly favors the 3D shape and its only difference is an alteration in the exchange-component of the functional form~\cite{hamadaVanWaalsDensity2014}. 
In Secs.~\ref{si-sec:au20_benchmarks_geom} and~\ref{si-sec:au20_benchmarks_zpe} of the Supporting Information, we show that neither the DFT model geometry nor zero-point energy effects (ZPE) significantly affect this energy ordering, providing confidence in these observations. 
Our predicted 3D-2D energy difference of the PBE functional is also in agreement with the study by Ricci \etal{}~\cite{ricciBondingTrendsDimensionality2006a} to be $\sim$1.2 eV, which studied similar 3D and 2D structures as this work with the closely related PW91 functional~\cite{perdewAtomsMoleculesSolids1992}.
In Sec.~\ref{si-sec:au20_benchmarks} of the Supporting Information, we identify another 2D structure that is also unstable relative to the 3D tetrahedral structure.
Ferrando \etal{}~\cite{ferrandoStructuresSmallAu2011a} have studied a further selection of 3D and 2D structures, reaching the same conclusion that the tetrahedral 3D shape is most stable. %

\section{Conclusions}

In summary, we introduce two approaches that now enable nanoclusters supported on oxide surfaces to be studied to a CCSD(T) level of accuracy.
First, with the SKZCAM protocol, we  obtain reliable CCSD(T)-quality simulations of small ($N\leq4$) coinage metal (Au$_N$, Ag$_N$ and Cu$_N$) nanoclusters adsorbed on MgO.
Second, the DFT+$\Delta$CC approach extends this accuracy to coinage metals up to $N=20$ and beyond.

We developed a dataset with the SKZCAM protocol to benchmark a broad set of DFT models for their performance in describing the metal-metal and metal-support interactions within oxide-supported nanoclusters.
While no DFT models could adequately describe metal-metal interactions, we found that PBE0-TS/HI could reach sub-chemical accuracy on the metal-support interactions.
The DFT+$\Delta$CC approach corrects for these metal-metal interaction errors and we demonstrate that PBE0-TS/HI+$\Delta$CC achieves near-chemical accuracy for coinage metal tetramers on MgO, as confirmed against the SKZCAM protocol.

The combination of these two reference methods has now allowed us to critically evaluate previous literature on oxide-supported nanoclusters. 
In particular, we showed with the SKZCAM protocol that there is a strong tendency for dispersion correction schemes (most notably the D3 dispersion) to overestimate the binding of coinage metal atoms on the MgO surface, particularly for Au.
This overbinding of the interaction between Au atoms and the MgO surface causes planar structures to be preferred, hence resulting in the anomalous predictions of a planar \ce{Au20} structure on the MgO system. 
With PBE0-TS/HI+$\Delta$CC, we show that the 3D tetrahedral structure is the most stable, in agreement with experimental evidence.

These methods, particularly PBE0-TS/HI+$\Delta$CC, open many exciting possibilities in the future. 
We can, for example, go beyond \ce{Au20} on MgO, where there is debate on the existence of open-cage structures~\cite{ferrandoSurfaceSupportedGoldCages2009} with novel catalytic properties. 
Additionally, our approach can also be used to study catalytic reactions on the nanoclusters; sizes beyond \ce{Au8} have shown promising catalytic activities~\cite{yoonChargingEffectsBonding2005c}. 
In these cases, a $\Delta$CC description of the catalytic molecules would be mandatory if we study their reactions on the surface, as DFT has well known deficiencies for reaction barriers~\cite{tealeDFTExchangeSharing2022}. 
Finally, we note that the DFT+$\Delta$CC approach, which only requires a CCSD(T) description of gas-phase nanoclusters, is highly amenable for combination with machine learning potentials trained on CCSD(T) data -- a burgeoning field~\cite{daruCoupledClusterMolecular2022,chenDataEfficientMachineLearning2023a}. 
This approach opens up the possibility of studying larger structures involving defects~\cite{yoonChargingEffectsBonding2005c} or incorporating dynamical properties, such as their fluxional character, which play a significant role in their catalytic activity~\cite{sunActiveSiteFluxional2021a}.

\section{Data Availability}

The Supporting Information is available free of charge at https://pubs.acs.org/doi/xx.xxxx. See the Supporting Information for a detailed compilation of the obtained results as well as further data and analysis to support the points made throughout the text. The input and output files associated with this study and all analysis can be found on GitHub at \href{https://github.com/benshi97/Data_Nanocluster_on_MgO}{benshi97/Data\_Nanocluster\_on\_MgO} or viewed (and analyzed) online on \href{https://colab.research.google.com/github/benshi97/Data_Nanocluster_on_MgO/blob/main/analyse.ipynb}{Colab}. The codes used to carry out this work are described and referenced in the Methods section and are available free-of-charge with the exception of VASP.

\subsection{Notes}
All authors declare no financial or non-financial competing interests.

\section{Acknowledgements}
The authors are grateful for: resources provided by the Cambridge Service for Data Driven Discovery (CSD3) operated by the University of Cambridge Research Computing Service (\href{www.csd3.cam.ac.uk}{www.csd3.cam.ac.uk}), provided by Dell EMC and Intel using Tier-2 funding from the Engineering and Physical Sciences Research Council (capital grant EP/P020259/1), and DiRAC funding from the Science and Technology Facilities Council (\href{www.dirac.ac.uk}{www.dirac.ac.uk}); 
the Cirrus UK National Tier-2 HPC Service at EPCC (\href{http://www.cirrus.ac.uk}{http://www.cirrus.ac.uk}) funded by the University of Edinburgh and EPSRC (EP/P020267/1); the Swiss National Supercomputing Centre (CSCS) under project ID s1052; 
the Korea Institute of Science and Technology Information (KISTI) for the Nurion cluster (KSC-2021-CRE-0542, KSC-2022-CRE-0115); 
and computational support from the UK national high performance computing service, ARCHER 2 obtained via the UKCP consortium and funded by EPSRC grant ref EP/P022561/1. 
BXS acknowledges support from the EPSRC Doctoral Training Partnership (EP/T517847/1).
AM acknowledges support from the European Union under the ``n-AQUA'' European Research Council project (Grant No.\ 101071937).
CWM acknowledges the support from the National Research Foundation of Korea (NRF) grant funded by the Korea government (MSIT) (No. NRF-2022R1C1C1010605).

\end{document}